\documentclass[useAMS,usenatbib,a4paper,preprint]{mn2e}
\usepackage{graphicx}
\usepackage{amsmath}
\usepackage{txfonts}
\usepackage{mathrsfs}

\def\aap{A\&A}
\def\apjl{ApJ}

\def\apjs{ApJS}

\newcommand{\be}{\begin{equation}}
\newcommand{\ee}{\end{equation}}
\newcommand{\bea}{\begin{eqnarray}}
\newcommand{\eea}{\end{eqnarray}}
\newcommand{\hb}{\mathrm{H}\beta}
\newcommand{\ha}{\mathrm{H}\alpha}

\title[HII Galaxy Hubble Diagram]{A Two-point Diagnostic for the HII Galaxy Hubble Diagram}
\author[Leaf \& Melia]{Kyle Leaf$^{1}$\thanks{kyleaf@email.arizona.edu} and
Fulvio Melia$^{2}$\thanks{John Woodruff Simpson Fellow. E-mail: fmelia@email.arizona.edu} \\
$^1$Department of Physics, The University of Arizona, AZ 85721, USA \\
$^2$Department of Physics, The Applied Math Program, and Department of Astronomy, 
The University of Arizona, AZ 85721, USA}

\begin{document}

\date{}

\pagerange{\pageref{firstpage}--\pageref{lastpage}} \pubyear{2017}

\maketitle

\label{firstpage}

\begin{abstract}
A previous analysis of starburst-dominated HII Galaxies and HII regions
has demonstrated a statistically significant preference for the
Friedmann-Robertson-Walker cosmology with zero active mass, known as the
$R_{\rm h}=ct$ universe, over $\Lambda$CDM and its related dark-matter
parametrizations. In this paper, we employ a 2-point diagnostic with
these data to present a complementary statistical comparison of $R_{\rm h}=ct$
with {\it Planck} $\Lambda$CDM. Our 2-point diagnostic compares---in a 
pairwise fashion---the difference between the distance modulus measured 
at two redshifts with that predicted by each cosmology. Our results 
support the conclusion drawn by a previous comparative analysis demonstrating 
that $R_{\rm h}=ct$ is statistically preferred over {\it Planck} $\Lambda$CDM. 
But we also find that the reported errors in the HII measurements may not 
be purely Gaussian, perhaps due to a partial contamination by non-Gaussian 
systematic effects. The use of HII Galaxies and HII regions as standard
candles may be improved even further with a better handling of the systematics
in these sources. 
\end{abstract}


\newpage
\section{Introduction}
Starbursts dominate the total luminosity of massive, compact galaxies known
as HIIGx. The closely related giant extragalactic HII~regions (GEHR) also
undergo massive bursts of star formation, but tend to be located predominantly
at the periphery of late-type galaxies. In both environments, the ionized
hydrogen is characterized by physically similar conditions (Melnick et al.
1987), producing optical spectra with strong Balmer $\ha$ and $\hb$ emission
lines that are indistinguishable between these two groups of sources (Searle
\& Sargent 1972; Bergeron 1977; Terlevich \& Melnick 1981; Kunth \& \"{O}stlin 2000).

Since both the number of ionizing photons and the turbulent velocity
of the gas in these objects increase as the starburst becomes more massive,
HIIGx and GEHR have been recognized as possible standard candles, a rather
exciting prospect given that the very high starburst luminosity facilitates
their detection up to a redshift $z\sim 3$ or higher (e.g., Melnick et al.
2000; Siegel et al. 2005). The exact cause of the correlation between
the luminosity $L(\hb)$ in $\hb$ and the ionized gas velocity dispersion
$\sigma$ is not yet fully understood, though an explanation may be found
in the fact that the gas dynamics is almost certainly dominated by the
gravitational potential of the ionizing star and its surrounding environment
(Terlevich \& Melnick 1981). These sources may therefore function as standard
candles because the scatter in the $L(\hb)$ versus $\sigma$ relation appears
to be small enough for HIIGx and GEHRs to probe the cosmic distance scale
independently of $z$ (Melnick et al. 1987; Melnick et al. 1988; Fuentes-Masip
et al. 2000; Melnick et al. 2000; Bosch et al. 2002; Telles 2003; Siegel et al.
2005; Bordalo \& Telles 2011; Plionis et al. 2011; Mania \& Ratra 2012;
Ch{\'a}vez et al. 2012, 2014; Terlevich et al. 2015).

Over the past several decades, HIIGx and GEHRs have been used to measure
the local Hubble constant $H_0$ (Melnick et al. 1988; Ch{\'a}vez et al.
2012), and to sample the expansion rate at intermediate redshifts (Melnick
et al. 2000; Siegel et al. 2005). More recently, Plionis et al. (2011) and
Terlevich et al. (2015) demonstrated that the $L(\hb)-\sigma$ correlation
is a viable high-$z$ tracer, and used a compilation of 156 combined sources,
including 24 GEHRs, 107 local HIIGx, and 25 high-$z$ HIIGx, to constrain
the parameters in $\Lambda$CDM, producing results consistent with Type Ia
SNe. Most recently, we (Wei et al. 2017)
extended this very promising work even further by demonstrating that
GEHRs and HIIGx may be utilized, not only to refine and confirm the
parameters in the standard model but---perhaps more importantly---to
compare and test the predictions of competing cosmologies, such as
$\Lambda$CDM and the $R_{\rm h}=ct$ universe (Melia 2003, 2007, 2013a,
2016, 2017a; Melia \& Abdelqader 2009; Melia \& Shevchuk 2012).

These two models have been examined critically using diverse sets
of data, including high-$z$ quasars (e.g., Kauffmann \& Haehnelt 2000;
Wyithe \& Loeb 2003; Melia 2013b, 2014; Melia \& McClintock 2015b), cosmic
chronometers (e.g., Jimenez \& Loeb 2002; Simon et al. 2005; Melia \& Maier
2013; Melia \& McClintock 2015a), gamma-ray bursts (e.g., Dai et al. 2004;
Ghirlanda et al. 2004; Wei et al. 2013), Type Ia supernovae (e.g., Perlmutter
et al. 1998; Riess et al. 1998; Schmidt et al. 1998; Melia 2012; Wei et al.
2015b), and Type Ic superluminous supernovae (e.g., Inserra \& Smart 2014;
Wei et al. 2015a). Their predictions have also been compared using the age
measurements of passively evolving galaxies (e.g., Alcaniz \& Lima 1999;
Lima \& Alcaniz 2000; Wei et al. 2015c). A more complete summary of these
comparisons, now based on over 20 different types of observation, may be
found in Table~1 of Melia (2017b).

The application of HIIGx and GEHRs as standard candles has provided one of
the more compelling outcomes of this comparative study involving $\Lambda$CDM
and $R_{\rm h}=ct$ (Wei et al. 2016). Using the combined sample of Ch\'avez
et al. (2014) and Terlevich et al. (2015), we constructed the Hubble diagram
extending to redshifts $z\sim 3$, beyond the current reach of Type Ia SNe,
and confirmed that the proposed correlation between $L(\hb)$ and $\sigma$
is a viable luminosity indicator in both models. This sample is already
large enough to demonstrate that $R_{\rm h}=ct$ is favored over $\Lambda$CDM
with a likelihood $\gtrsim 99\%$ versus only $\lesssim 1\%$, corresponding
to a confidence level approaching $3\sigma$.

These results, however, come with two important caveats, which partially
motivate the complementary approach we are taking in this paper. Not surprisingly,
the cosmological parameters are most sensitive to the high-$z$ data, so the
constraints resulting from this work are heavily weighted by the high-$z$
sample of only 25 HIIGx. Given how sensitive the results are to the sub-sample
of high-$z$ HIIGx data, one would want to increase the significance of this
analysis by increasing the number of HIIGx-related measurements. Indeed,
with the K-band Multi Object Spectrograph at the Very Large Telescope, a
larger sample of high-$z$ HIIGx high-quality measurements may be available
soon (Terlevich et al. 2015).

The second caveat attached to the analysis of Wei et al. (2017) is that we do not yet
have a full grasp of the systematic uncertainties in the $L(\hb)-\sigma$ correlation;
these no doubt impact the use of HIIGx as cosmological probes. They include the
burst size, its age, the oxygen abundance of HIIGx, and the internal extinction
correction (Ch{\'a}vez et al. 2016). An example of a non-ignorable systematic
uncertainty arises from the fact that the $L(H\beta)-\sigma$ relation correlates the
ionizing flux from massive stars with random velocities in the potential well created by
all the stars and the surrounding gas. Thus, any systematic variation in the initial
mass function would alter the mass-luminosity ratio, and therefore also the
zero point and slope of the relation (Ch\'avez et al. 2014).

In spite of the fact that the high-$z$ sample of HIIGx is still relatively
small, we can nonetheless further test the previous results by probing this
compilation more deeply (than has been attempted before) using a two-point
diagnostic, $\Delta\mu(z_i,z_j)$, defined in Equation~(9) below. Quite generally,
two-point diagnostics such as this differ from parametric fitting approaches in
several distinct ways. They facilitate the comparative analysis of measurements
in a pairwise fashion. One may use them with $n$ measurements of a particular
variable to generate $n(n-1)/2$ comparisons for each pair of data. 
The benefits are twofold:
(1) one can test how well each pair of data fits the models, and (2) assess
how closely the published error bars fit a normal distribution, thereby
providing some indication of possible contamination by correlated systematic
uncertainties. Zheng et al. (2016) recently used such an approach
to conclude that the stated errors in cosmic chronometer data are strongly
non-Gaussian, suggesting that the quoted measurement uncertainties are
almost certainly not based exclusively on statistical randomness (see also
Leaf \& Melia 2017).

As we shall see, the diagnostic $\Delta$$\mu$$(z_i,z_j)$ is expected to be
zero if the model being tested is the correct cosmology. To allow for
possible non-Gaussianity in the published errors, we shall use both
weighted-mean and median statistics to determine the degree to which
each model's distribution of $\Delta$$\mu$$(z_i,z_j)$ values is consistent
with this null result. So while Wei et al. (2016) optimized the overall
$\Lambda$CDM and $R_{\rm h}=ct$ parametric fits to the HII~galaxy Hubble
diagram, here we will test the consistency of each fit with individual
pairs of data. We will begin with a brief description of the data in \S~2,
and then define and apply the diagnostic $\Delta$$\mu$$(z_i,z_j)$ in \S~3.
The outcome of our analysis will be discussed in \S~4, followed by our
conclusions in \S~5.

\section{Observational Data and Methodology}
We base our analysis on the methodology described in Ch\'avez et al.
(2012, 2014) and Terlevich et al. (2015), using their total sample of 156 
sources, including 107 local HII~galaxies, 24 giant extragalactic HII~regions, 
and 25 high-$z$ HII~galaxies. The correlation between the emission-line 
luminosity and the ionized gas velocity dispersion may be written as 
(Ch\'avez et al. 2012; Ch\'avez et al. 2014; Terlevich et al. 2015)
\begin{equation}
\log L(\hb)=\alpha \log \sigma(\hb)+\kappa\;,
\end{equation}
where $\alpha$ is the slope and the constant $\kappa$ represents
the logarithmic luminosity at $\log \sigma(\hb)=0$. As noted, previous
applications of this relation have produced a very small scatter in the
correlation for $L(\hb)$, making it a viable luminosity indicator for
cosmology. But one cannot completely avoid its cosmology dependence
because the $\hb$ luminosity is calculated using the
expression
\begin{equation}
L(\hb) = 4 \pi D_L^2(z) F(\hb)\;,
\end{equation}
where $D_L$ is the model-dependent luminosity distance at redshift $z$
and $F(\mathrm{H}\beta)$ is the reddening corrected $\hb$ flux. 

From Equation~(1), we may then obtain the distance modulus of an HII~galaxy according to
\begin{equation}
\mu_{\rm obs}=2.5\left[\kappa +\alpha \log \sigma(\hb) - \log F(\hb)\right]-100.2 \;,
\end{equation}
with an associated error
\begin{equation}\label{sigma}
\sigma_{\mu_{\rm obs}}=2.5\left[\left(\alpha \sigma_{\log \sigma}\right)^{2}+
\left(\sigma_{\log F}\right)^{2}\right]^{1/2} \;,
\end{equation}
in terms of $\sigma_{\log \sigma}$ and $\sigma_{\log F}$, these being the $1\sigma$ uncertainties
in $\log \sigma(\hb)$ and $\log F(\hb)$, respectively. This is to be compared with
the theoretical distance modulus
\begin{equation}
\mu_{\rm th}\equiv5 \log\left[\frac{D_{L}(z)}{\rm Mpc}\right]+25\;,
\end{equation}
as a function of the cosmology-dependent luminosity distance $D_L$.

\begin{table*}
\begin{center}
\caption{Parameters optimized via maximization of the likelihood function}
\begin{tabular}{lccccc}
&&&& \\
\hline\hline
&&& \\
Model& $\alpha$ & $\delta$ & $\Omega_{\rm m}$ & $\Omega_{\rm de}$ & $w_{\rm de}$ \\
&&&& \\
\hline
&&&& \\
$R_{\rm h}=ct$ &$ 4.78_{-0.09}^{+0.07}$ & $32.01_{-0.30}^{+0.32}$ & -- & -- & -- \\
&&&& \\
{\it Planck} $\Lambda$CDM & $4.86_{-0.08}^{+0.08}$ & $32.27_{-0.31}^{+0.22}$ & $0.3089$ & $1.0-\Omega_{\rm m}$ & $-1$ \\
&&&& \\
$\Lambda$CDM & $4.86_{-0.10}^{+0.09}$ & $32.27_{-0.36}^{+0.34}$ & $0.32_{-0.06}^{+0.09}$ & $1.0-\Omega_{\rm m}$ & $-1$ \\
&&&& \\
\hline\hline
\end{tabular}
\end{center}
\end{table*}

In $\Lambda$CDM, the luminosity distance may be written
\begin{eqnarray}\label{DL_LCDM}
\qquad D_{L}^{\Lambda {\rm CDM}}(z) &=& {c\over
H_{0}}\,{(1+z)\over\sqrt{\mid\Omega_{k}\mid}}\times {\rm sinn}\Biggl\{\mid
\Omega_{k}\mid^{1/2}\times\cr 
\null &\null&\hskip-0.6in
\int_{0}^{z}{dz\over\sqrt{\Omega_{\rm m}(1+z)^{3}+\Omega_{k}(1+z)^{2}+
\Omega_{\rm de}(1+z)^{3(1+w_{\rm de})}}}\Biggr\}\;,
\end{eqnarray}
where $p_{\rm de}=w_{\rm de}\rho_{\rm de}$ is the dark-energy equation
of state; radiation is ignored in the local Universe. Also, $\Omega_i\equiv
\rho_i/\rho_c$, for matter (m), radiation (r) and dark energy (de), while 
$\Omega_{k}=1-\Omega_{\rm m}-\Omega_{\rm de}$ incorporates the spatial curvature of the
Universe, and sinn is $\sinh$ when $\Omega_{k}>0$ and $\sin$ when $\Omega_{k}<0$.
Today's critical density is $\rho_c\equiv 3c^2 H_0^2/8\pi G$.
Since we are here assuming a flat Universe, (i.e., $\Omega_{k}=0$), the right side
of this equation becomes $(1+z)c/H_{0}$ times the integral. For the
$R_{\rm h}=ct$ cosmology (Melia 2003, 2007, 2013a, 2016a, 2016b; Melia \& Abdelqader 2009;
Melia \& Shevchuk 2012), the luminosity distance is given by the much simpler expression
\begin{equation}\label{DL_Rh}
D_L^{R_{\rm h}=ct}(z)={c\over H_0}(1+z)\ln(1+z)\;.
\end{equation}

Here we follow Wei et al's. (2015b,2016) approach and circumvent circularity 
issues by optimizing the coefficients $\alpha$ and $\kappa$ individually for each 
model, via maximization of the likelihood function. With this approach, $H_{0}$ and 
$\kappa$ are not independent of each other; one may vary either $H_{0}$ or $\kappa$, 
but not both. For the purpose of maximizing the likelihood function, it is therefore 
useful to define a combined parameter,
\begin{equation}
\delta\equiv -2.5\kappa-5\log H_0+125.2\;,
\end{equation}
where $\delta$ is the ``$H_0$-free" logarithmic luminosity and the Hubble constant
$H_0$ is in units of km $\rm s^{-1}$ $\rm Mpc^{-1}$. The constants $\alpha$ and $\delta$ are
statistical ``nuisance" parameters, analogous to the adjustable coefficients characterizing
the lightcurve in Type Ia SNe. The best-fit parameters obtained in this fashion are 
shown in Table~1, for three models we will compare:
{\it Planck} $\Lambda$CDM (Planck Collaboration 2016), $\Lambda$CDM with a re-optimized 
matter density $\Omega_{\rm m}$, and the $R_{\rm h}=ct$ universe.

A quick inspection of Equations~(3) and (5) shows that the two-point diagnostic
\begin{eqnarray}
\qquad{\Delta}{\mu}{(z_i,z_j)}&\equiv&\dfrac{-{\delta}+2.5{\alpha}\log\sigma_i-2.5\log F_i}{5\log
\left[\dfrac{D_L(z_i)}{1\;{\rm Mpc}}\right]}-\cr
\null&\null&\qquad\dfrac{-{\delta}+2.5{\alpha}\log\sigma_j-
2.5\log F_j}{5\log\left[\dfrac{D_L(z_j)}{1\;{\rm Mpc}}\right]}
\end{eqnarray}
is expected to be zero for any pair of HII~data at redshifts $z_i$ and $z_j$ if
the cosmology used to calculate $D_L$ is correct. As one can see, the value of
$H_0$ does not affect this constraint and is absorbed into the optimized coefficient
$\delta$. For the sake of normalizing the various quantities, however, we simply use
the {\it Planck} value $67.74$ km s$^{-1}$ Mpc$^{-1}$ throughout this analysis.

Notice in passing that $\alpha$ and $\delta$ are similar between the different 
cosmologies, varying between them by $\lesssim 4\%$, i.e., well within $1\sigma$. 
Thus, since $H_0$ is also not a factor in $\Delta\mu(z_i,z_j)$, Equation~(9) represents 
a powerful diagnostic for comparing the viability of different models. The application 
of this 2-point diagnostic will be described in the next section.

Finally, to improve the statistics even further, we have removed seventeen points 
(including one GEHR source at z=000001) from our complete sample whose measurement 
places them more than $3\sigma$ away from the best fit curves. We have also chosen 
to remove the other GEHR source at z=0.00001. While this point is only $~2\sigma$ from 
the best fit curve, it is the lowest-redshift measurement in the catalog, 
which, by the nature of 2-point diagnostics, causes it to drastically alter the statistical results.
These anomalous points are identical for all three models, so their removal does 
not bias either of them. The final reduced sample therefore contains 138 measurements that are used 
to determine the best fits reported in Table 1. The eighteen eliminated sources are the two 
GEHRs at z=0.00001, and J162152+151855, J132347-013252, J211527-075951,
J002339-094848, J094000+203122, J142342+225728, J094252+354725,\break
J094254+340411, J001647-104742, J002425+140410, J103509+094516, J003218+150014, 
J105032+153806, WISP173-205, J084000+180531, and Q2343-BM133.

\section{Application of the Two-point Diagnostic}
As discussed in more detail in Leaf \& Melia (2017), the use of two-point diagnostics necessitates
special care when analyzing the statistics they produce. First, the weighted mean of all
$n(n-1)/2$ $\Delta$$\mu(z_i,z_j)$ values may be calculated using the expression
\begin{equation}
\mu=\dfrac{{\Sigma_{i=1}^{n-1}{\Sigma}_{j=i+1}^{n}{\Delta}\mu(z_i,z_j)/
{\sigma}_{{\Delta}\mu_{ij}}^2}}{{{\Sigma}_{i=1}^{n-1}\Sigma_{j=i+1}^{n}1/{\sigma}_{{\Delta\mu_{i,j}}}^2}}\;,
\end{equation}
in which ${\sigma}_{{\Delta\mu_{i,j}}}$ is the error for a single application of Equation~(9),
found using standard error propagation. The error in the mean, however, must be calculated
by carefully considering the correlation introduced from the repeated use of individual
points in different pairs. For this purpose, we rewrite the weighted mean in the equivalent form
\begin{equation}
\mu=\dfrac{\Sigma_{i=1}^{n}{\beta_i}M(z_i)}{{\Sigma}_{i=1}^{n-1}
\Sigma_{j=i+1}^{n}1/{\sigma}_{{\Delta\mu_{i,j}}}^2}\;,
\end{equation}
with each $\beta$ value given by the expression
\begin{equation}
\beta_i=\Sigma_{j=1}^{i-1}\dfrac{1}{\sigma_{\Delta\mu_{i,j}}^2}-\Sigma_{k=1+i}^{N}\dfrac{1}
{\sigma_{\Delta\mu_{i,k}}^2}\;.
\end{equation}
In addition, we have defined the quantity
\begin{equation}
M(z_i)=\dfrac{-{\delta}+2.5{\alpha}\log\sigma_i-2.5\log F_i}{5\log
\left[\dfrac{D_L(z_i)}{1\;{\rm Mpc}}\right]}\;.
\end{equation}
With the values of $\beta$ thus calculated, the variance then follows and is given as
\begin{equation}
{\Delta}{\sigma^2}_{\rm w.m.}=\dfrac{\Sigma_{i=1}^{n}{\beta^2_i}
\sigma_{M}^2(z_i)}{({\Sigma}_{i=1}^{n-1}\Sigma_{j=i+1}^{n}1/{\sigma}_{\Delta\mu_{i,j}}^2)^2}\;.
\end{equation}
Knowing the standard deviation of the mean, we now have a measure of the consistency
of the measurements with a given model. In the case of the $\Delta\mu$ diagnostic,
we expect the weighted mean to be statistically consistent with zero if the applied
model is the correct cosmology. Note that we do not introduce the errors in the fitted
parameters in this analysis. This is due to the error affecting both halves of the two-point
diagnostics in a very similar manner. That is, if the value of $\alpha$ is slightly too low, it would
have the effect of reducing both `single-points', the net effect of which ends up being statistically
insignificant.

When non-Gaussian errors are suspected, however, such situations motivate the use
of `median statistics,' pioneered by Gott et al. (2001), in which error
propagation is neither required nor assumed. This approach takes advantage of
the fact that for any measurement based on some distribution function,
there is a $50\%$ chance of it being above the true median of the
underlying distribution, without any need to know its form. Thus,
for $N$ ranked measurements, the true median has a probability
\begin{equation}
P_i=\dfrac{2^{-N}N!}{i!(N-i)!}
\end{equation}
(i.e., the binomial distribution) of being found between measurements
$i$ and $i+1$. One can use this to construct confidence regions about
the median of the data, analogous to a standard deviation in Gaussian
statistics, and assign to them a formal probability of finding the
true median of the underlying distribution. However, it would be incorrect to
apply this to, say all $n(n-1)/2$ diagnostics at once, for the 
same reasons noted in Leaf \& Melia (2017).
The fact that each measurement contributes to N-1 diagnostics means that the
data are correlated; as a result, a single measurement can move the median farther
than in the case where the 2-point values are truly randomly distributed.

We propose a remedy that takes advantage of the binomial properties of the
median, but instead of considering all the diagnostics simultaneously, we 
construct a random sub-sample, in which each realization of the diagnostic is 
used exactly once, except for the one that was omitted. Therefore none of the 
diagnostic values is used more than once, completely avoiding any possible 
correlation. Following this, we record the median of the diagnostics of this 
uncorrelated sample, as well as the standard deviation of the realization. 
Next, we generate a large number (here, 1 million) of these realizations, 
and report the overall median of all the individual medians in Table 2. 

\begin{table*}
\begin{center}
\caption{Statistical analysis of the 2-point diagnostic $\Delta\mu(z_i,z_j)$}
\begin{tabular}{lccccccc}
&&&&&& \\
\hline\hline
&&&&&& \\
Model& {\rm Weighted mean} & $1\sigma$ {\rm Error} & $|$Mean$|$ / $\sigma$ &$|N_\sigma|<1$ & {\rm Median} & Std. Dev. of the Median 
& $|$Median$|$ / Std. Dev.\\
&&&&&& \\
\hline
&&&&&& \\
$R_{\rm h}=ct$ &$-0.00242$ & $0.00218$& $1.11$ & $51.3\%$ & $-0.00425$ & $0.00336 $ & $1.26$ \\
&&&&& \\
{\it Planck} $\Lambda$CDM & $-0.00340$ & $0.00221$& $1.54$ & $52.3\%$ & $-0.00483$ & $0.00363$ & $1.33$\\
&&&&& \\
$\Lambda$CDM & $-0.00330$ & $0.00220$& $1.50$ & $52.2\%$ & $-0.00476$ & $0.00342$ & $1.39$\\
&&&&&& \\
\hline\hline
\end{tabular}
\end{center}
\end{table*}

\begin{figure}
\begin{center}
\includegraphics[width=1.0\linewidth]{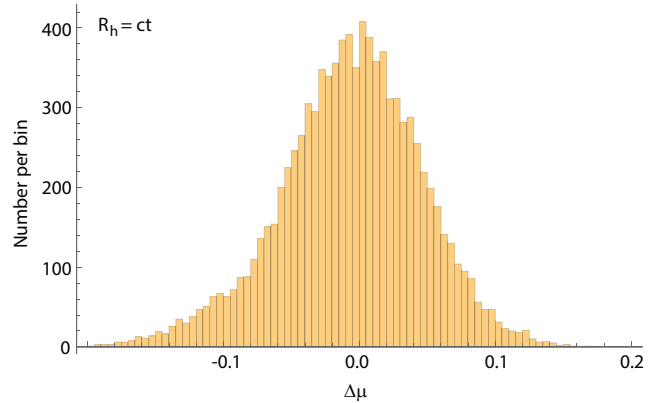}
\end{center}
\caption{Unweighted histogram of all 9,453 $\Delta\mu$ diagnostic values for
the $R_{\rm h}=ct$ universe (see Eq.~9). The $y$-axis gives the number of diagnostic 
values per bin.}
\end{figure}

\begin{figure}
\begin{center}
\includegraphics[width=1.0\linewidth]{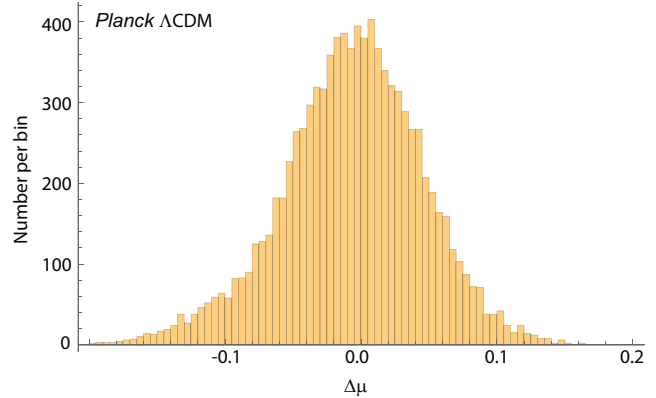}
\end{center}
\caption{Same as fig.~1, except now for {\it Planck} $\Lambda$CDM.}
\end{figure}

\begin{figure}
\begin{center}
\includegraphics[width=1.0\linewidth]{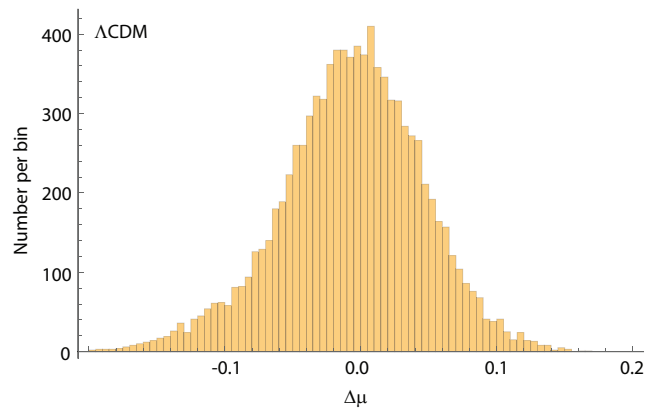}
\end{center}
\caption{Same as fig.~1, except now for $\Lambda$CDM with a re-optimized
value of $\Omega_{\rm m}$ (see Table~1).}
\end{figure}

In table 2, we also report the standard deviation of the median. This value is different from 
the overall standard deviation of the set of all 1 million medians. It is fundamentally related
to the error in the mean of any set of data, in that it is some distinct factor smaller than the 
standard deviation of the data, dependent on the size of the data set. However, the exact relationship 
that exists between the standard deviation of the medians and the number of sources used to determine 
the median of all the realizations is not empirically known. 

In order to address this deficiency, we have have used the following approach, based on Monte-Carlo 
simulations with mock data to find this relationship to reasonable accuracy. We construct a mock 
data set by drawing at random from some probability distribution function, with the same number 
(i.e., 138) of points as in the real data set. Then, we construct a random set of 2-point 
diagnostics following the same method used with the real data. We record the median and standard 
eviation of the realization, repeating this process a sufficiently large number of times 
(say, $20,000$). Then, we repeat the process with a new random set of mock data drawn from the same 
distribution, and repeat this 5,000 times. Next, we determine the standard deviation of the set of 
5,000 medians, as well as the mean of the 5,000 standard deviations. Finally, we compare the actual 
standard deviation of the median of all realizations with the mean of the standard deviations of each 
realization. We run this simulation with three different probability density functions---a normal 
distribution, a skew normal distribution with shape parameter $\alpha$=4, and a flat distribution 
over an interval. In all three cases the relationship between the standard deviation of the median 
and the mean standard deviation of each realization is found to be statistically consistent, 
and apparently dependent only on the number of sources chosen. 

For a sample of 138, the multiplicative factor is 1.822, always yielding a standard deviation 
of the medians smaller than the mean of the standard deviations by this factor. 
The values reported in Table 2 for the standard deviation 
of the median are therefore determined by taking the standard deviations of the million medians
and dividing them by the corresponding factor. While this does technically include an implicit 
assumption that all data are sampled from a single underlying statistical distribution, 
we argue that by focusing on the median of these (instead of the mean), and the fact that there 
must certainly exist a single true cosmological model, this assumption is reasonable.

The two-point diagnostic we have introduced in Equation~(9) is expected to be zero for the
correct cosmology. The degree by which a given model's median is consistent with zero
is therefore a measure of its consistency with the observations. We discuss the results
of this analysis in the next section.

\section{Discussion}
In Table~2 and figs.~1-6 we report the results of both our weighted-mean and median statistical 
analyses, described in \S\S~2 and 3 above. One of the principal benefits of 2-point diagnostics constructed 
with regard to redshift ordering lies not only in determining how well a set of data fits a model, as 
revealed, e.g., with the use of information criteria but, also in providing insight into whether or not 
the low-$z$ sources are consistent with the same model as that preferred by the higher-$z$ sources.

Our complete sample of 138 sources constitutes the original 156 minus the
18 outliers, as detailed in \S~2. As one can see from Table 1, the optimized value of
$\alpha$ is about $4.8$ in every case, statistically consistent with the results of
previous analyses by Ch\'avez et al. (2012, 2014), Terlevich et al. (2015), and
Wei et al. (2016). For these 138 measurements, we constructed for each model the 
9453 unique 2-point diagnostics and calculated the weighted mean and corresponding
1-$\sigma$ error based on the reported uncertainties (see figs.~1-3 for the 
complete unweighted histograms). For the $R_h=ct$ universe (fig.~4), the weighted mean 
is found to be consistent with zero at about $1\sigma$. There is mild tension for
{\it Planck }$\Lambda$CDM (fig.~5) and the best-fit $\Lambda$CDM cosmology (fig.~6), 
however, in that the weighted mean is inconsistent with zero at about $1.5\sigma$ 
(compare the entries in columns 2, 3 and 4 of Table 2). Perhaps more importantly, 
fewer than the expected $68.3\%$ of the diagnostics lie within $1\sigma$ of the 
weighted mean (column 5) for all three models, implying that the reported errors 
are probably not purely Gaussian and that there may be an additional source of error
not accounted for in this analysis. 

It is therefore helpful to circumvent this possible non-Gaussianity by also analyzing 
the 2-point diagnostics using median statistics, as described above. With this approach, 
the three models show a similar inconsistency with a zero median (columns 6, 7
and 8 of Table 2),
with a negative value in every case, roughly $1.3\sigma$ different from zero. The fact
that both the weighted mean and the median are negative for all the models suggests that 
the luminosity distance at low-$z$ is generally greater than that predicted by these
cosmologies, or that it is smaller than expected at high-$z$. The implication is that 
either (i) none of the models are completely correct, or (ii) there may be some systematic
problems with the data at high-$z$ or (more likely) at low-$z$. Thus, while a discrepancy 
smaller than $2\sigma$ may not be definitive, it nonetheless motivates further analysis 
involving a possible contamination by non-Gaussian systematic errors.

Along these lines, we point out that some authors have speculated on the possibility
that a local ``Hubble bubble" (Shi 1997; Keenan et al. 2013; Romano 2016) might
be influencing the local dynamics within a distance $\sim 300$ Mpc (i.e., $z\lesssim 
0.07$). If true, such a fluctuation might lead to anomalous velocities within this 
region, causing the nearby expansion to deviate somewhat from a pure Hubble flow. 
This effect could be the reason we are seeing a slight negative bias for the
weighted mean and median of the 2-point diagnostic for every model, since nearby
velocities would be slightly larger than Hubble, implying larger than expected
luminosity distances at redshifts smaller than $\sim 0.07$. In addition, the
existence of local peculiar velocities would imply that the errors associated 
with low-$z$ measurements should be bigger than quoted, increasing the number
of 2-point diagnostics that fall within $1\sigma$ of the expected dispersion,
possibly `filling' the distributions in figs.~1-3 sufficiently to produce 
entries in column 5 of Table~2 closer to the value ($\sim 68.3\%$) expected
of a true Gaussian distribution.

\begin{figure}
\begin{center}
\includegraphics[width=1.0\linewidth]{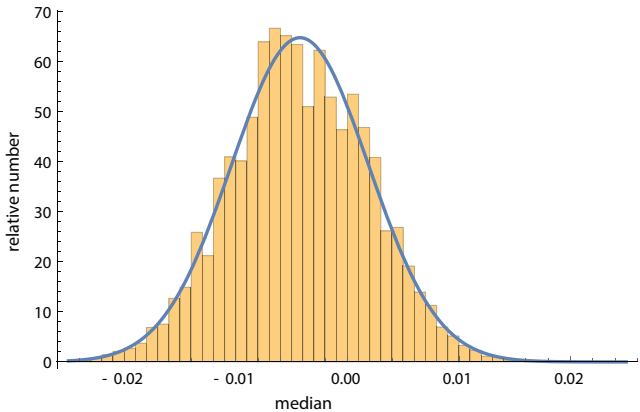}
\end{center}
\caption{Histogram of the medians found in one million random realizations of the 2-point
diagnostic for the $R_{\rm h}=ct$ universe. The $y$-axis denotes
the number of times ($\times 1000$) that the median of a realization falls within the range
given on the $x$-axis.}
\end{figure}

\begin{figure}
\begin{center}
\includegraphics[width=1.0\linewidth]{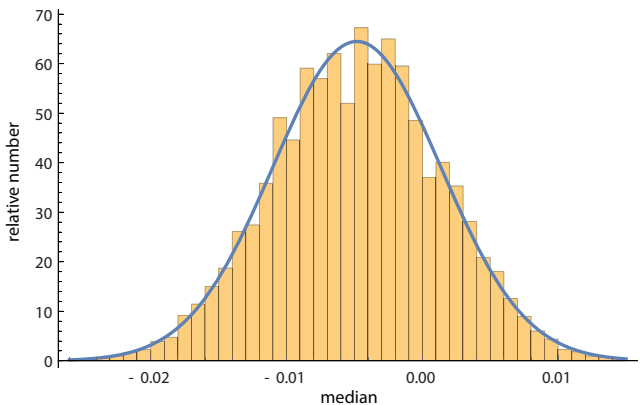}
\end{center}
\caption{Same as figure~4, except for {\it Planck} $\Lambda$CDM.}
\end{figure}

\begin{figure}
\begin{center}
\includegraphics[width=1.0\linewidth]{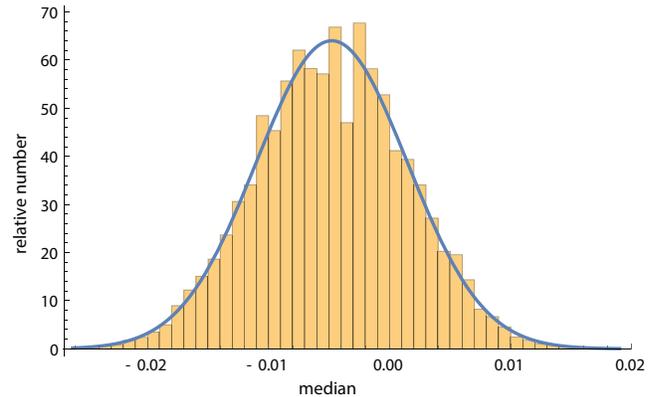}
\end{center}
\caption{Same as figure~4, except for $\Lambda$CDM with a re-optimized
$\Omega_{\rm m}$, as indicated in Table~1.}
\end{figure}

\section{Conclusions}
The totality of the results shown in Tables~1 and 2, and illustrated
in figs.~1-6, supports the use of HIIGx and GEHR sources as standard candles
for cosmological testing, though the analysis based on 2-point diagnostics has 
probed the measurement errors in greater detail than was possible solely via 
parametric fits to the data, the subject of our previous paper on this
subject (Wei et al. 2016).

In this paper, we have proposed a new 2-point diagnostic for analyzing HIIGx and 
GEHR data with the inclusion of median statistics, which circumvents the need for 
assuming Gaussian errors in the measurements. This approach may be used alongside, 
and compared, with the better understood weighted mean method. We have shown that
these two types of analysis give generally consistent results, insofar as the HII
data are concerned. Broadly speaking, one of the principal conclusions of this
analysis is that employing the entire compilation of HIIGx and GEHR sources 
(with the exception of several outliers) produces slight tension between the 
cosmological parameters favoured by the data at low and high redshifts. We believe 
this is circumstantial evidence in support of the proposal by Shi (1997), Keenan 
et al. (2013) and Romano (2016) of a dynamical influence due to a local Hubble 
bubble extending out to $z\sim 0.07$, which produces local peculiar velocities 
comparable to those in the Hubble flow at low redshifts.

Nonetheless, probing the HIIGx and GEHR data with 2-point diagnostics has not 
changed the essential conclusions drawn by Wei et al. (2016), whose cosmological 
tests based on these sources favoured the $R_{\rm h}=ct$ model over $\Lambda$CDM. 
Our comparison using the HII sample has shown that $R_{\rm h}=ct$ is favoured over 
both {\it Planck} $\Lambda$CDM and $\Lambda$CDM with a variable $\Omega_{\rm m}$, 
at least when viewed in terms of weigthed mean statistics. The caveat, however,
is that an approach based on median statistics produces less differentiation
between the three models.

In addition, we have found in all cases that our 2-point diagnostic with the weighted 
mean approach yields fewer values within individual $1\sigma$ error regions than the
$68.3\%$ required of a true Gaussian distribution. This may be an indication that the 
reported errors are not purely statistical, which may happen, e.g., when the 
uncertainties are contaminated by systematic effects, including at least a partially 
non-Gaussian component, or when there is an additional source of uncertainty, other 
than what we considered in this analysis.

\section*{Acknowledgments}
We are very grateful to the anonymous referee for providing
valuable insights and suggested improvements to the manuscript.
FM is supported by Chinese Academy of Sciences Visiting Professorships for
Senior International Scientists under grant 2012T1J0011, and the Chinese
State Administration of Foreign Experts Affairs under grant GDJ20120491013.

\label{lastpage}

\end{document}